%% file: paper_la.tex
\documentclass[twocolumn,showpacs,aps,prd,superscriptaddress]{revtex4}

\usepackage{graphicx}
\usepackage{dcolumn}
\usepackage{amsmath}
\usepackage{epsfig}

\input pubboard/babarsym
\input definitions

\newcommand{\BABARPubYear}    {03}
\newcommand{\BABARPubNumber}  {022}

\newcommand{\SLACPubNumber} {10162}

\def\figurebox#1#2#3{%
    \def\arg{#3}%
    \ifx\arg\empty
    {\hfill\vbox{\hsize#2\hrule\hbox to #2{\vrule\hfill\vbox to #1{\hsize#2\vfill}\vrule}\hrule}\hfill}%
    \else
    {\hfill\epsfbox{#3}\hfill}%
    \fi}

\begin{document}

\preprint{\babar-PUB-\BABARPubYear/\BABARPubNumber} 
\preprint{SLAC-PUB-\SLACPubNumber} 

\begin{flushleft}
\babar-PUB-\BABARPubYear/\BABARPubNumber\\
SLAC-PUB-\SLACPubNumber\\
\end{flushleft}

\title{
\vskip 10mm
{\large \bf
Measurement of $\sin 2\beta$ with Hadronic and Previously Unused Muonic 
$J/\psi$ Decays} 
}

\input authorsl

\date{\today}

\begin{abstract}
We report a measurement of the $CP$-violation parameter $\sin 2\beta$ with
$B^0 \rightarrow J/\psi K^0_S$ decays in which the $J/\psi$ decays to hadrons 
or to muons that do not satisfy our standard identification criteria.
With a sample of 88 million \BB\ events collected by the \babar\ detector at 
the PEP-II asymmetric-energy $B$ factory at SLAC, we reconstruct $100 \pm 17$ such events, with $J/\psi\to \pi^+ \pi^- \pi^0$ being the most
 prevalent, and
 measure $\sin 2\beta = 1.56 \pm 0.42 ({\rm stat.}) \pm 0.21 ({\rm syst.})$. 
\end{abstract}


\maketitle

Measurement of \CP\ violation in the $B$-meson system, particularly in
$b \rightarrow c\overline{c}s$ transitions, has been a primary goal of the \babar\ experiment. In the 
Standard 
Model, these decays exhibit a \CP\ asymmetry that is proportional 
to $\sin2\beta$, where $\beta$ is defined as
$\arg[-V_{cd}^{ }V_{cb}^*/V_{td}^{ }V_{tb}^*]$,  with $V_{ij}$ the  
elements of the CKM matrix~\cite{ref:CKM}.  The current world average 
value of \stwob\ is $0.731 \pm 0.056$~\cite{ref:PDG}, 
with the $B$ factories (\babar\ at SLAC and Belle at KEK)
providing the most precise measurements~\cite{ref:BaBarsin2b, ref:Bellesin2b}.
The dominant decay mode in these measurements
is $B^0 \rightarrow J/\psi K^0_S$, where only leptonic decays of the $J/\psi$ 
are considered.
Leptonic decay modes have the advantage of low backgrounds, but account for only 12\% of $J/\psi$ 
decays~\cite{ref:PDG}.  Since the current measurements of $\sin 2\beta$ 
are statistically
limited, in this article we extend the measurement through the use of hadronic 
$J/\psi$ decays, as well as previously unused muonic decays.

 At the $B$ factories,  $B^0$ mesons are produced via 
$\epem\to\FourS\to\B^0\bar{B}^0$.  For $B^0$ mesons produced in this 
manner and decaying to the \CP\ eigenstate $\jpsi K^0_S$, 
\stwob\ appears as the amplitude of a time-dependent \CP\ asymmetry.  The 
Standard 
Model 
predicts the decay rate 
$$
f_\pm(\Delta t) = {e^{-|\Delta t|/\tau_{B^0}} \over 4\tau_{B^0}}[1 \pm \stwob\sin(\Delta m_d\Delta t)],
$$
where the plus (minus) sign indicates that the other, "tagging", $B^0$ meson 
in the event decays as a $B^0(\bar{B}^0)$, $\Delta t$  is the decay time of 
the \CP-eigenstate $B^0$ meson minus the decay time of the tagging 
$B^0$ meson, 
$\tau_{B^0}$ is the $B^0$ lifetime, and $\Delta m_d$ is the 
mass difference between the two mass-eigenstate neutral $B$ mesons 
($\Delta m_d$ is also the $B^0-\bar{B}^0$ oscillation frequency). The 
time-dependent \CP\ asymmetry is:
$$
A_{\CP} \equiv { f_+(\Delta t) - f_-(\Delta t) \over f_+(\Delta t) + f_-(\Delta t)} = \stwob\sin(\Delta m_d\Delta t).
$$
Measurement of $A_{\CP}$ requires that a sample of
$B^0$ mesons decaying to  $\jpsi K^0_S$ be reconstructed, that the
flavor of the other $B^0$ meson in the event be 
determined, and that $\Delta t$ be measured.  

A sample of $88 \pm 1$ million $B\bar{B}$ events recorded by the
 \babar\ detector~\cite{ref:NIMpaper} was used in this analysis.
The innermost component of \babar\ is a five-layer
double-sided silicon microstrip vertex detector with $90^{\circ}$ stereo 
angle, allowing precise reconstruction of the location of the $B^0$ decay vertices along 
the beam 
direction. Since the $\FourS$ is boosted along the beam direction, the 
difference in position between the $B^0$ decay vertices in this direction allows one to measure
$\Delta t$.  The primary tracking device is a 40-layer drift chamber 
 operated with a helium-based gas mixture to  minimize multiple scattering.  The drift chamber is surrounded by a 
Cherenkov
particle identification device, and a CsI(Tl) 
calorimeter. All of the above detectors reside in a 1.5 T field generated by a 
superconducting solenoid.  The flux is returned via layers of steel interleaved
with active detectors for the identification of muons and detection of neutral
hadrons.

Two types of Monte Carlo simulated events are used in the analysis.  One,
called ``full MC'', 
consists
of events that are generated according to the known physics of 
$B\bar{B}$ and continuum
production, passed through a detailed
 model of the detector response~\cite{ref:GEANT},
 and reconstructed in the same manner
as the data sample.  The second, called ``parametrized MC'', consists of events for 
which the
relevant parameters are randomly generated according to the distributions 
observed in data or in detailed simulations. For any study where an accurate 
model of the physics or detector response is required, full MC is used.  
Parametrized MC, which can be generated more quickly than full MC, is only
used to explore the statistical properties of the extraction of $\sin 2 \beta$.

While many \jpsi\ decays to exclusive hadronic final states have been 
observed~\cite{ref:PDG}, the sum of their measured branching fractions is less than
20\%.   To allow for the possibility of observing a signal in previously 
unmeasured decay modes, we take an inclusive approach in the first stage of 
event selection.  Charged
tracks are assigned either the electron, muon, pion, kaon, or proton mass 
based on
particle identification information, and candidates for 
$\pi^0 \rightarrow \gamma\gamma$ and $\eta \rightarrow \gamma\gamma$ or 
$\pi^+\pi^-\pi^0$ are formed.  All neutral combinations of up to six tracks 
and neutral mesons are considered (a maximum of two neutral mesons
is allowed), and those consistent with baryon number conservation, strangeness
conservation, and Bose symmetry, and having invariant mass $m_{\jpsi}$ in the 
range $2.80 - 3.20 \gevcc$, are retained for further analysis.  Decay modes
of the type $\jpsi\to KK\pi$ are excluded to ensure that the selected 
sample is 
independent of the sample used in \babar's previous measurement of
\stwob~\cite{ref:BaBarsin2b}, which included $B^0\to\etac\KS$ events with $\etac\to KK\pi$.

We form  \KS\ candidates from 
a pair of oppositely-charged tracks that have invariant mass between 489 and
507 \mev\ and a vertex displaced by at least 1 mm from the $\jpsi$ candidate's 
vertex.
The selected $\jpsi$ and \KS\ candidates are combined to form $B^0$ candidates.  Two kinematic variables are used to 
isolate the $B$ meson signal: the difference \DeltaE\ between the energy of 
the  
reconstructed $B$ candidate and the beam energy in the center-of-mass frame,
and the beam-energy substituted mass \mes $\equiv 
\sqrt{E_{\rm beam}^{*2}-p_B^{*2}}$,
where $p_B^*$ is the momentum of the reconstructed $B$ and $E_{\rm beam}^*$
is the beam energy, both in the center-of-mass frame.
The small variations of $E_{\rm beam}^{*}$ within the data sample
are taken into account when calculating \mes.  Signal events will have
\DeltaE\ close to 0 and values of \mes\ close to the $B^0$ meson mass.
Candidates are required to have $\mes > 5.20 \gevcc$ and $|\DeltaE| < 55 \mev$
 if the \jpsi\ decays entirely to charged particles, 
and $ < 105 \mev$ if the decay includes one or more neutral hadrons.
The $\DeltaE$ selection accepts candidates within $3\sigma$ of the distribution
observed in simulated signal events.   The resolution in $\mes$ is 3\mev, so the selection admits a large region at low $\mes$ in addition to the region populated by signal candidates.  Inclusion of this sideband region allows the magnitude of the combinatoric background to be measured.  
   
Backgrounds arise both from continuum $q\bar{q}$ production and from $B$
 meson decays to other modes.  The continuum events tend to have a 
two-jet 
topology, in contrast to the more spherically symmetric $B\bar{B}$ events.
A set of 18 variables (described in~\cite{ref:etacks}) that are sensitive to 
this difference are combined in
a Fisher discriminant ${\cal F}$, which is defined to have an average value of
1 for signal and -1 for continuum events.  The weight of each variable in the 
discriminant is
calculated by maximizing the separation between a sample of data taken below
the $B\bar{B}$ threshold (and thus composed entirely of continuum $q\bar{q}$
 events) 
and a sample of simulated signal events. We place progressively tighter
requirements on ${\cal F}$ as the candidate $\jpsi$ decay multiplicity increases:
for two-body decays we require ${\cal F} > -1.14$, for three-body decays we 
require ${\cal F} > -0.70$, and for higher-multiplicity decays we require 
 ${\cal F} > -0.37$.

For three-body $\jpsi$ decays, which have a larger combinatoric background
than two-body decays, additional separation between signal and background
is attained by considering
 the angle $\theta_d$ between the normal to the 
plane in which the momenta of the $\jpsi$ daughter particles lie and the 
\KS\ direction in the $\jpsi$ rest frame.  Conservation of 
angular momentum requires  this variable to be distributed as $\cos^2\theta_d$ for $J/\psi$ decays to three pseudoscalars (the most common type of three-body
decays),
while it is 
uniformly distributed for $B\bar{B}$ backgrounds and peaks at $\cos\theta_d = 0$ for continuum $q\bar{q}$ backgrounds.  We require candidates to
have $|\cos\theta_d| > 0.55$.  The selection in $\cos\theta_d$ and ${\cal F}$ 
was
chosen to maximize $S/\sqrt{S+B}$, where $S$ is the expected signal and $B$ the
expected background.

There are two classes of $B\bar{B}$ backgrounds.  The first consists of 
candidates formed from a subset of a given $B$ meson's decay
products, or from a combination of decay products from the two $B$ mesons in 
the event.  This background and the continuum $q\bar{q}$ background are 
henceforth referred to as ``combinatoric backgrounds''.  They have a linearly 
falling distribution in \DeltaE,
and their distribution in \mes\ may be parametrized by an empirical
phase-space distribution~\cite{ARGUS_bkgd} (henceforth referred to as the 
ARGUS function):
\begin{eqnarray}
A(\mes;m_0,c_{\rm arg})& \propto &\mes\sqrt{1-(\mes/m_0)^2} \times \nonumber \\
             &         & \exp(c_{\rm arg}(1-(\mes/m_0)^2)), \nonumber
\end{eqnarray}
where $m_0$ is a cutoff mass set to 5.291 \gev (a typical center-of-mass beam 
energy) and $c_{\rm arg}$ is a fitted parameter.

The second class of $ B\bar{B}$ background consists of $B$ mesons that decay to a 
topology also allowed for $J/\psi\KS$, but without a 
$\jpsi$ in the intermediate state.  These ``peaking'' backgrounds are 
dominated by $B$ decays that have a
charmed meson in the intermediate state, so we remove any candidates for which a 
$D$ or $D^*$ meson within 2$\sigma$ of the nominal mass can be formed from the 
final-state hadrons. Since these backgrounds arise from fully 
reconstructed $B^0$ mesons, they have the same distribution in \mes\ and 
\DeltaE\ as the signal.

Since the branching fractions for many of the modes that contribute to 
the peaking backgrounds are not well-measured, we must extract the peaking
background magnitude from the data.  We do this by performing a two-dimensional
 unbinned maximum likelihood fit to the \mes\ and $m_{\jpsi}$ distributions. 
The likelihood function used is:
\begin{eqnarray}
 L & = & (n_{\rm comb}A(\mes;m_0,c_{\rm arg}) + (n_{\rm sig}+n^0_{\rm peak})G(\mes)) \times \nonumber \\
  & &  ((n_{\rm comb} + n^0_{\rm peak})C(m_{\jpsi}; p_1, p_2) + n_{\rm sig}G(m_{\jpsi})), \nonumber
\end{eqnarray}
where $n_{\rm comb}$ is the fitted combinatoric background, $n^0_{\rm peak}$ is
 the fitted peaking background, $ n_{\rm sig}$ is the fitted signal, $A$ is 
a normalized ARGUS function, $G$ are normalized Gaussians, and $C$ is a 
normalized second-order Chebyshev polynomial with parameters $p_i$. 
The mean and width of $G(\mes)$ are fixed to the values observed in 
high-statistics hadronic $B$-decay samples, and the mean and width of 
$G(m_{\jpsi})$ are fixed to the values observed in our  $\jpsi\to\mumu$ 
sample for two-body
decay modes, and to the values observed in full MC events for 
higher-multiplicity modes. 
The photon-energy resolution in the simulated events is
degraded to match that observed in data.  The additional smearing required 
is 3\% of the measured photon energy for photons below 100 \mev, and decreases with increasing photon energy (no additional smearing is required for photons above 1 \gev).

The $J/\psi$ decay modes for which the measured signal magnitude is less than 
its statistical uncertainty are removed from the analysis.  
The surviving modes, and their 
contribution to the signal, are listed in Table~\ref{tab:fitByMode-Ks}.  Note 
that no modes including an $\eta$ meson are observed, and also that no decays
with a multiplicity of greater than three are visible above background.

\begin{table}
\begin{center}
\begin {tabular}{lccc}
\hline
\jpsi\ decay mode &  Signal & Peaking Bkg. & Comb. Bkg. \\ \hline
\pipi & $ 28 \pm 8 $ & $84 \pm 17$ & $206 \pm 12$ \\
$K^+K^-$ & $ 5 \pm 3 $ & $-1 \pm 6$ & $42 \pm 5$  \\
$p\bar{p}$ & $ 6 \pm 3 $ & $1 \pm 6$ & $34 \pm 5$ \\ \hline
Total $h^+h^-$ & $40 \pm 9$ & $86 \pm 19$ & $279 \pm 13$ \\
After final selection & $28 \pm 8$ & $13 \pm 3$ & $15 \pm 3$ \\ \hline
\pipi\piz & $ 58 \pm 17 $ & $104 \pm 29$ & $652 \pm 23$ \\
$p\bar{p}\piz$ & $ 11 \pm 6 $ & $9 \pm 9$ & $77 \pm 7$ \\ \hline
Total $h^+h^-\piz$ & $69 \pm 18$ & $113 \pm 30$ & $716 \pm 22$ \\
After final selection & $72 \pm 13$ & $19 \pm 5$ &  $74 \pm 8 $ \\ \hline
\end{tabular}
 
\end{center}
\caption{Observed \BztoJpsiKS\ signal and background.  The combinatoric
backgrounds reported are the integral of the fitted ARGUS function in the
region $\mes > 5.27 \gevcc$. Except for the rows labelled ``After final
selection'', the numbers are measured
prior to application of the final selection criteria on $m_{\jpsi}$ 
and \DeltaE.  All uncertainties are statistical only.}
\label{tab:fitByMode-Ks}
\end{table}        

The observation of 28 candidates in the $J/\psi\to\pipi$ channel is
inconsistent with our expectation of observing about one event given the known branching fraction of 
$(1.47 \pm 0.23) \times 10^{-4}$ \cite{ref:PDG} for this mode.  We interpret 
the excess candidates as $J/\psi\to\mumu$ decays in which both muons fail the
standard muon selection criteria.  Studies using simulated events with 
muon identification efficiencies measured in data confirm
that the observed signal magnitude is consistent with the $J/\psi\to\mumu$ 
hypothesis.  Since these events do measure \stwob, and are independent of the events used in our previous measurements~\cite{ref:BaBarsin2b}, we retain them for this analysis. 

After $n^0_{\rm peak}$ is determined, 
the following final selection criteria are imposed
to improve the purity of the sample:  We recalculate 
$\DeltaE$ with the \jpsi\ candidate constrained to the nominal 
mass, and define the result as $\Delta E_c$.  The resolution in 
$\DeltaE_c$ is 11 \mev\ for two-body \jpsi\ decay candidates, and 12 \mev\ for three-body candidates.  For two-body \jpsi\ decay
candidates we require $3.06 <  m_{\jpsi} < 3.12\gevcc$ and $|\Delta E_c| < 33\mev$, and for three-body  \jpsi\ decay
candidates we require $3.05 <  m_{\jpsi} < 3.15\gevcc$ and $|\Delta E_c| < 35\mev$.  

\begin{figure}
\begin{center}
\epsfig{file= 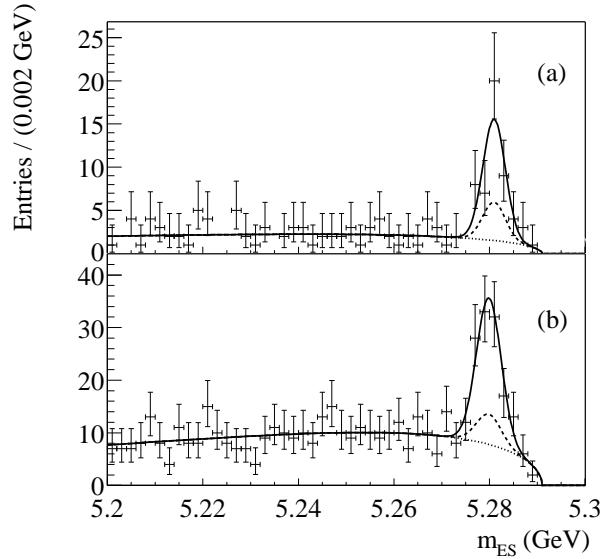,width=\columnwidth} \\
\end{center}
 \caption{
\mes\ distributions for candidates for \BztoJpsiKS\ with the \jpsi\ decaying to
(a) two and (b) three particles.  The dotted line represents the fitted 
combinatoric
background distribution. The dashed line represents the total background
distribution,
while the solid line represents the signal plus background distribution.
\label{fig:finalFits-Ks}}
\end{figure}

The 
efficiency of this selection for peaking
 backgrounds ($\varepsilon_{\rm peak}$) is 
estimated using full MC events. We define $\varepsilon_{\rm peak}$ as the 
ratio of the  area of the fitted Gaussian in \mes\ after the final selection
 to the area 
before the final selection.  For two-body decay candidates $\varepsilon_{\rm peak} = 0.15 \pm 0.01 ({\rm stat.})$ and for three-body decay candidates
 $\varepsilon_{\rm peak} = 0.17 \pm 0.02 ({\rm stat.})$.  An unbinned maximum likelihood fit to
the sum of a Gaussian distribution and an ARGUS function is 
performed on the \mes\ distributions of the surviving
candidates.  The integral of the ARGUS function measures the 
combinatoric background, while the integral of the Gaussian measures the sum of
the signal and peaking background.  Subtracting $n_{\rm peak} \equiv 
\varepsilon_{\rm peak}n^0_{\rm peak}$ from the latter provides an estimate of the signal.  The \mes\ distributions are shown in Fig.~\ref{fig:finalFits-Ks}, and the signal and background magnitudes in the final sample are
reported in Table~\ref{tab:fitByMode-Ks}.

Once the sample of \BztoJpsiKS\ candidates has been isolated, the extraction of
\stwob\ proceeds in the same manner as for \babar's other recent 
measurements~\cite{ref:BaBarsin2b}.  Information from the final-state 
particles recoiling against the $J/\psi\KS$ meson candidate
is used to determine whether the other $B$ meson in the event was a
$B^0$ or $\bar{B}^0$ at the time of its decay.  This is referred to as the 
flavor ``tag''.  The variables used for tagging
include the charge of any high-momentum identified electron or muon, 
the charge
of any identified kaon, and the charge of a slow pion consistent with 
arising from $D^*$ meson decay.  The efficiency $\varepsilon$ 
and mistag
rate $w$ are measured using the data
as described below, and reported in Ref.~\cite{ref:BaBarsin2b};
the overall figure of merit for the flavor-tagging performance,
$\varepsilon(1-2w)^2$, is $ (28.1 \pm 0.7)$\%.

\begin{figure}
\epsfig{ file = 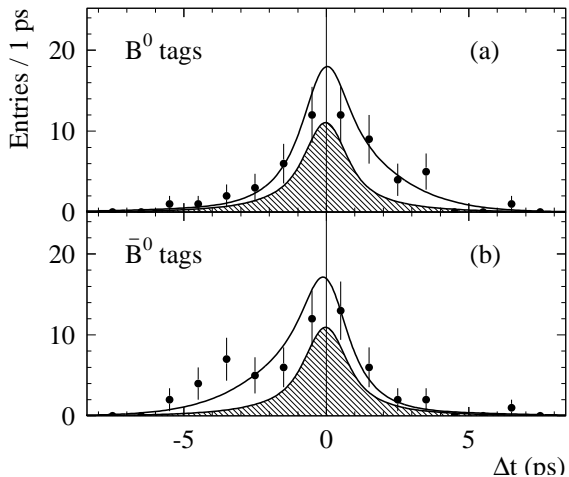, width=\columnwidth}
\caption { $\Delta t$ values observed in the \BztoJpsiKS\ candidates.
The plots show the distribution for events in which the recoiling
$B$ meson is tagged as (a) $B^0$ and (b) $\bar{B}^0$.  In each plot the 
solid line represents the
result of the maximum likelihood fit, and the shaded area the contribution of
background.
\label{fig:jpsixksDT}}
\end{figure}

\begin{figure}
\epsfig{ file = 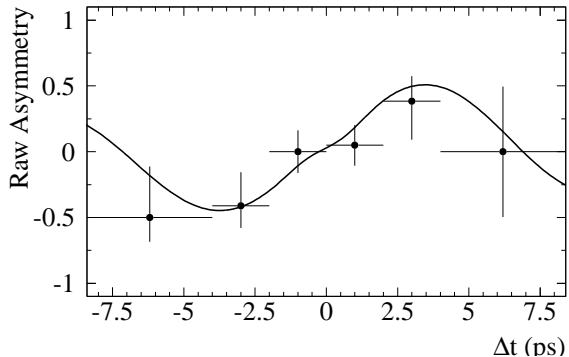, width=\columnwidth}
\caption {  $\Delta t$ asymmetry observed before correction for backgrounds and
mistag probability in \BztoJpsiKS\ candidates, with
best-fit asymmetry displayed.
\label{fig:jpsixksAsym}}
\end{figure}

The extraction of \stwob\ is done using an unbinned maximum likelihood fit
to the $\Delta t$ distribution of the candidate events, where the assumed
functional form is $f_\pm(\Delta t)$ convolved with the resolution of the
$\Delta t$ measurement, with the mistag probability taken into account.
The input to the fit consists of both the signal sample and a large sample 
of fully reconstructed 
$B^0$ decays to $D^{(*)+}\pi^-$, $D^{(*)+}\rho^-$, $D^{(*)+}a^-$ and 
$J/\psi K^{0*}$ with $K^{0*} \rightarrow K^+\pi^-$.   
The $B^0$ flavor is known for these modes, so this sample 
constrains a set of parameters describing the
flavor-tagging performance and vertex resolution,  The simultaneous fit 
takes into account any 
 correlations between these parameters and the value of \stwob.
  The result is:
$$
\stwob = 1.56 \pm 0.42 \hbox{ (stat.)}
$$
The $\Delta t$ distribution for flavor-tagged signal events is shown in
Fig.~\ref{fig:jpsixksDT}, and the \CP\ asymmetry observed
before correction for backgrounds and mistag probability is shown 
in Fig.~\ref{fig:jpsixksAsym}.
In each case a projection of the best-fit model is superimposed.

As a cross-check, the analysis was repeated using a sample of
$\BchtoJpsiKch$ events selected in a manner analogous to the \CP\ sample, 
and with the same \jpsi\ decay modes considered.  
This sample yields an apparent \stwob\ of $-0.13 \pm 0.20 \hbox{ (stat.)}$,
consistent with the expected null result.

Systematic uncertainties arise from several sources.  In performing the fit
for \stwob\ it is assumed that the background has no \CP\ asymmetry.  Since 
some of the background is composed of real $B^0$ mesons this may not be true.
Fitting for \stwob\ on a sample composed of candidates in the $m_{\jpsi}$ or 
$\Delta E_c$ sidebands yields $0.18 \pm 0.46$.  The signal sample
is then refit with the \CP\ asymmetry of the peaking background
fixed to the $\pm1\sigma$ limits of the measured asymmetry,  and the 
observed variation of $\pm 0.15$ in \stwob\ is taken as a systematic 
uncertainty.

The next most significant systematic uncertainty arises from the estimation 
of the
background magnitudes.  When the \stwob\ fit is performed, the parameter
$c_{\arg}$ of the ARGUS
distribution describing the combinatoric background is fixed to the central
value determined from fitting the \mes\ distribution.  The \stwob\ fit is 
repeated with this value fixed to its $\pm 1\sigma$ limits, and the observed 
variation in \stwob\ of $\pm 0.13$ is taken as a systematic uncertainty.

The uncertainty on the peaking background arises from several sources, the 
largest of which is the statistical uncertainty on $n^0_{\rm peak}$.  
The next most
significant source is uncertainty in $\varepsilon_{\rm peak}$. We estimate the
 magnitude of this uncertainty by observing the variations in 
$\varepsilon_{\rm peak}$ among samples of different simulated $B^0$ decay 
modes.  In addition, one could define $\varepsilon_{\rm peak}$ as the 
efficiency for any candidate with $\mes\ > 5.27 \gev$  to pass the 
final selection, rather than defining it as the ratio of fitted Gaussian
areas.   
We take the difference between the two definitions as a systematic.  The 
estimate of $n^0_{\rm peak}$ is also subject to uncertainty in the 
distribution of peaking
backgrounds in $m_{\jpsi}$, which is modeled as a second-order Chebyshev
polynomial. The variation in $n^0_{\rm peak}$ when the order is changed by
$\pm 1$ is propagated to the systematic uncertainty.  The accuracy of the fit used to extract the signal is verified using background-only 
samples, such as data recorded below the $B\bar{B}$ threshold or samples
of candidates reconstructed in modes not accessible to the $J/\psi$.
No statistically-significant signal yields are reported in fits to these
samples.
We assign the largest artificial signal yield consistent with these tests as 
a systematic uncertainty.  The final source is
the uncertainty on the resolution of the $J/\psi$ peak (which is held fixed in
the fit that determines $n^0_{\rm peak}$).  Variation of this assumed width 
between
values observed in different decay modes yields a variation in $n^0_{\rm peak}$.  
The sum in quadrature of all these effects totals 25\% of the magnitude of 
$n_{\rm peak}$.  Repeating the fit on many samples of parametrized MC events, each of which
has the same size and background as the sample observed in data, shows that the
variation in \stwob\ resulting from a 25\% uncertainty in the peaking 
background is $\pm 0.07$.

There are potentially differences in the flavor-tagging performance and
vertex resolution between events with hadronic $J/\psi$ decays and the other
fully-reconstructed $B$ decays used to measure these parameters.
Performing a \stwob\ fit to a large sample of full MC signal events with
$J/\psi\to\pi^+\pi^-\pi^0$ with the flavor tagging
and vertex resolution fixed to the measured values 
yields a result consistent with the generated value.  The 
statistical uncertainty of the result ($\pm 0.04$) is taken as a systematic 
uncertainty.

Another systematic uncertainty arises from events in which one or more of the 
final state 
particles assigned to the reconstructed $B^0$ in fact originated from the other 
$B^0$ in the event.  The fraction of such events is negligible for two-body 
\jpsi\ decays, and about 5\% for three-body decays.  Performing \stwob\ fits 
on full MC samples with and without including the incorrectly reconstructed 
candidates yields a variation of $\pm 0.01$ in \stwob.

Finally we take into account all the sources of systematic uncertainty
  that apply to \babar's previous measurements of \stwob~\cite{ref:BaBarsin2b}, except
  for those specific to the $B^0 \rightarrow \jpsi\KL$ mode, that have not already
  been specifically addressed here.  These uncertainties primarily arise from limits on our
understanding of flavor-tagging and vertex reconstruction performance, and 
yield a variation of $\pm 0.03$ in \stwob.
 
The systematic uncertainties are summarized in
Table~\ref{tab:CPsyst}.  The sum in quadrature of all contributions is 
$0.21$.

\begin{table}
\begin{center}
\begin {tabular}{lc}
\hline
Source & Uncertainty \\ \hline
Peaking background CP & 0.15 \\
Combinatoric background magnitude & 0.13 \\
Peaking background magnitude & 0.07 \\
Tagging and vertexing differences & 0.04 \\
Common to leptonic modes & 0.03 \\
Misreconstructed signal & 0.01 \\ \hline
Total                   &  0.21 \\
\hline
\end{tabular}
\end{center}
\caption{Summary of systematic uncertainties on the measurement of \stwob.
\label{tab:CPsyst}}
\end{table}

The value of \stwob\ reported in this analysis is higher than the world 
average value of $0.731 \pm 0.056$. 
To estimate the consistency of this result with the world average, 
10,000 parametrized MC samples 
with the same signal and background magnitudes as observed in the data were
generated with a true \stwob\ of 0.731.  To simulate the systematic uncertainty
in this analysis and the total uncertainty on the world average, a
random number with Gaussian distribution and 
 $\sigma = 0.22$ is added to the \stwob\ result for each sample.  Of 
the 10,000 samples, 629 fluctuated
to a value of 1.56 or greater, indicating that the probability of such a
fluctuation is 6.3\%.

In summary, we have extended \babar's previous \stwob\ measurement
by including \jpsi\KS\ modes where the \jpsi\ decays to hadronic
final states.  The  result is
$$
\stwob = 1.56 \pm 0.42 \hbox{ (stat.)} \pm 0.21  \hbox{ (syst.).}
$$

Although we searched for
many hadronic \jpsi\ decay modes, signals were observed only in modes that have been
previously seen~\cite{ref:PDG}.  Further, only in hadron multiplicities of two
and three was it possible to observe a signal above background.  Extending the
analysis to the $\chi_c$ and $\psi(2S)$ mass regions does not yield additional
significant signals, nor is an $\eta_c$ signal observed after elimination
of $KK\pi$ modes.
 
\input pubboard/acknowledgements

\vskip12pt 
\hbox{$\ \ \ \ \ \ \ \ \ \ \ \ \ \ \ \ \ \ \ \ \ \ \ \ ${\LARGE{\bf---------------}}}

\end{document}

%% file: definitions.tex
\def\jpsi {\ensuremath{J/\psi}}

\def\BztoJpsiKS {\ensuremath{\Bz \to \jpsi \KS}}
\def\BchtoJpsiKch {\ensuremath{\Bpm \to \jpsi \Kpm}}

%% file: authorsl.tex
%
\author{B.~Aubert}
\author{R.~Barate}
\author{D.~Boutigny}
\author{J.-M.~Gaillard}
\author{A.~Hicheur}
\author{Y.~Karyotakis}
\author{J.~P.~Lees}
\author{P.~Robbe}
\author{V.~Tisserand}
\author{A.~Zghiche}
\affiliation{Laboratoire de Physique des Particules, F-74941 Annecy-le-Vieux, France }
% ...
\collaboration{The \babar\ Collaboration}
\noaffiliation

%% file: pubboard/acknowledgements.tex
We are grateful for the 
extraordinary contributions of our \pep2\ colleagues in
achieving the excellent luminosity and machine conditions
that have made this work possible.
The collaborating institutions wish to thank 
SLAC for its support and the kind hospitality extended to them. 
This work is supported by the
US Department of Energy
and National Science Foundation, the
Natural Sciences and Engineering Research Council (Canada),
Institute of High Energy Physics (China), the
Commissariat \`a l'Energie Atomique and
Institut National de Physique Nucl\'eaire et de Physique des Particules
(France), the
Bundesministerium f\"ur Bildung und Forschung
(Germany), the
Istituto Nazionale di Fisica Nucleare (Italy),
the Research Council of Norway, the
Ministry of Science and Technology of the Russian Federation, and the
Particle Physics and Astronomy Research Council (United Kingdom). 
Individuals have received support from the Swiss 
National Science Foundation, the A. P. Sloan Foundation, 
the Research Corporation,
and the Alexander von Humboldt Foundation.